\documentclass{sig-alternate}
\usepackage{url}
\usepackage{graphicx}
\usepackage{caption}
\usepackage{subcaption}
\usepackage{wrapfig}
\usepackage{multirow}
\usepackage{comment}
\usepackage{flushend}

\makeatletter
\def\@copyrightspace{\relax}
\makeatother

\begin{document}

%


\title{{\huge {\em ex uno pluria:}}\\ The Service-Infrastructure Cycle, Ossification, \\ and the Fragmentation of the Internet
}

%
%
%
%
%

\numberofauthors{1} 
%
\author{
%
%
\alignauthor
Mostafa H. Ammar 
\\
       \affaddr{School of Computer Science}\\
       \affaddr{Georgia Institute of Technology}\\
       \affaddr{Atlanta, GA}\\
       \email{ammar@cc.gatech.edu}
}


\maketitle
\begin{abstract}

In this article I will first argue that a {\em Service-Infrastructure Cycle} is fundamental to networking evolution. Networks are built to accommodate certain services at an expected scale. New applications and/or a significant increase in scale require a rethinking of network mechanisms which results in new deployments. Four decades-worth of iterations of this process have yielded the Internet as we know it today, a {\em common and shared} global networking infrastructure that delivers almost all services. 
I will further argue, using brief historical case studies, that success of network mechanism deployments often hinges on whether or not mechanism evolution follows the iterations of this Cycle.  
Many have observed that this network, the Internet, has become ossified and unable to change in response to new demands. In other words, after decades of operation, the Service-Infrastructure Cycle has become stuck. However, novel service requirements and scale increases continue to exert significant pressure on this ossified infrastructure.  The result, I will conjecture, will be a fragmentation, the beginnings of which are evident today, that will ultimately fundamentally change the character of the network infrastructure. By ushering in a ManyNets world, this fragmentation will lubricate the Service-Infrastructure Cycle so that it can continue to govern the evolution of networking.
I conclude this article with a brief discussion of the possible implications of this emerging ManyNets world on networking research.

\end{abstract}

\section{Introduction}

\begin{figure}[t]
\vspace{-0.3in}
  \centering
    \includegraphics[width=0.5\textwidth]{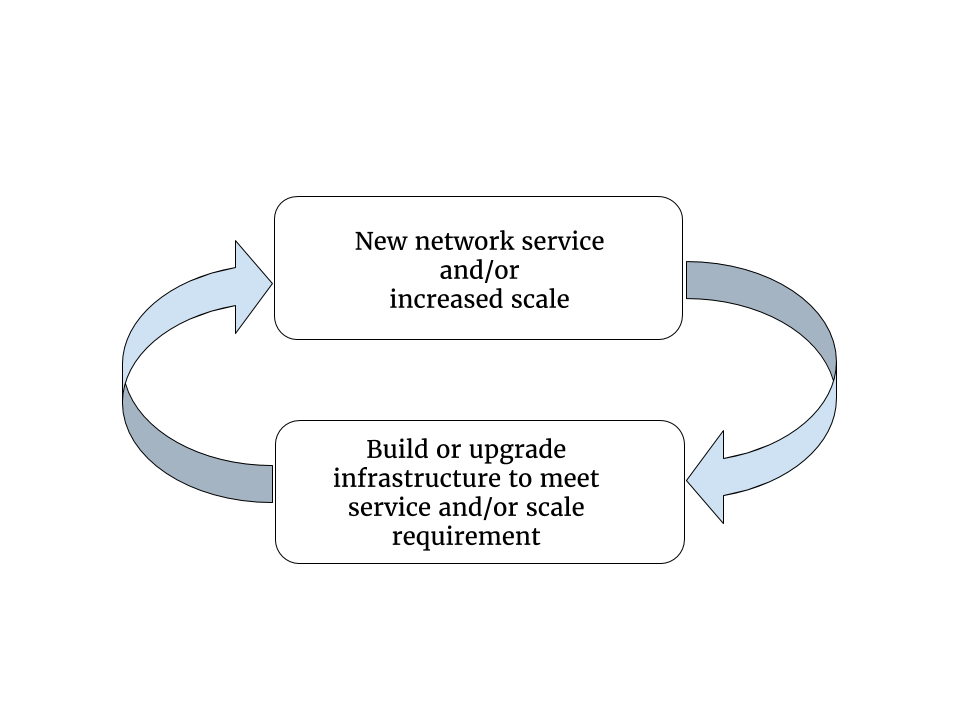}
 \vskip -20pt
    \caption{The Service-Infrastructure Cycle}
  \label{cycle}
\end{figure}

{\em Necessity is the mother of invention}. This has indeed been the case in the development of the global data networking infrastructure. In networking parlance one can restate the proverb as ``Service requirements motivate infrastructure deployment." With the ``necessity" that motivates networking  continuously evolving,  the ``invention" has never been a one-time innovation. Rather, a
{\em Service-Infrastructure Cycle} (or simply {\em the Cycle}) as shown in Figure \ref{cycle} has been the framework that governed networking evolution. 


In this framework, progress in {\em deployed} (as opposed to experimental or proposed) networking infrastructure is a continuous cycling. At any given point in time the networking infrastructure is designed and works well to support the popular uses and applications of the time. Network mechanisms and resources are also adapted to the current expected scale: the number of connected users, their network workload and their performance expectations. 

At any point in time, the network infrastructure, typically, has some flexibility to support new applications and its resources are provisioned with an anticipation of growth. As time goes on, however, new applications and/or a significant increase in scale of network usage stresses the network infrastructure in ways that require a rethinking of current mechanisms or a significant step increase in network resources. That, in addition to technological advances, results in new deployments that satisfy the changed service requirements.

Consider, for example, that in the early days data networking was built to enable remote access to mainframe computers \cite{leiner2009brief}. The system's design never anticipated email, which quickly emerged as the dominant use of the network. 
The evolution of the network continued to respond to new applications as file transfer, the web, content distribution, electronic commerce and video streaming challenged the design and deployment of the networking infrastructure. 

Another force driving changes in networks is scale. As the uses of the network multiplied the number of attached end points increased and the networking technology had to evolve to accommodate the scale. Examples of this over the last four decades, to name just a few, include the development of the Domain Name Systems (DNS) to provide large scale address resolution service, the hierarchical inter-domain structure of the current Internet, the adoption of classless IPv4 addressing, and the development and increasing deployment of IPv6.

{\em The Service-Infrastructure Cycle worked well to give us the Internet until it got stuck in the early 2000s.} In 2005 Anderson et al \cite{anderson2005overcoming} complained about
``the inability [of the Internet architecture] to adapt to new pressures and requirements." In related work around the same time Turner et al \cite{turner2005diversifying}
argued that the Internet infrastructure suffers from {\em ossification}. This was defined as a situation where ``alterations to the Internet architecture that address its fundamental deficiencies or enable new services have been restricted to incremental changes." It was further stated that this ossification ``stifles innovation and the adoption of disruptive technology." 
Despite new service demands or increased scale, the Cycle has become stuck and the network infrastructure is no longer amenable to modification in meaningful ways because it has become ossified. 
Ossification and the reasons for it have received significant attention in the literature \cite{turner2005diversifying,anderson2005overcoming}. To summarize, the scale of the Internet and the multiple authorities that govern its operation, make it difficult to agree on required changes and  to deploy new features.

In this article I consider the question of whether and how the data networking infrastructure can continue its evolution to respond to new demands despite the ossification barrier.
To this end, I will first argue that the Service-Infrastructure Cycle is fundamental to networking evolution.  Historically, success of network mechanism deployments often hinges on whether or not mechanism evolution follows the Cycle. In fact, anticipating the Cycle, i.e., trying to do too much too soon, can contribute to deployment failure.  In support of this argument, I present in Section \ref{ossification} two case studies on the evolution of unicast and multicast routing. The aim is to showcase deployment success when the mechanism evolution was allowed to follow the Service-Infrastructure Cycle, and failure when developments do not follow the iterative nature of the Cycle.

Next, in Section \ref{manynets}, I will start with the observation that four decades-worth of iterations of the Service-Infrastructure Cycle have yielded a {\em common and shared} global network that has become ossified. I will argue, however, that novel service requirements and scale increases continue to exert significant pressure on this ossified infrastructure, seriously hindering the operation of the Cycle.  The result, I will conjecture, will be a fragmentation, the beginnings of which are evident today, that will ultimately fundamentally change the character of the network infrastructure. This fragmentation will lubricate the Service-Infrastructure Cycle and help it get unstuck so that it can continue to govern the evolution of networking.

A few caveats: First, it is not my intention to (re)write networking or Internet history. There has been many excellent efforts in that regard  (see for example \cite{leiner2009brief} and \cite{hafner1998wizards}). Rather, my goal is to add structure to this history in an effort to understand it better. Second, my attempt to add structure and recognize patterns will leave large uncovered gaps in time and technology. I aim to get this type of discussion started and look forward as others contribute to improve on this humble start.

\section{A Tale of Two Routings}

\label {ossification}

To understand the workings of the Service-Infrastructure Cycle let us briefly trace the contrasting histories of unicast and multicast routing. I will argue that successful unicast routing deployment progressed until its Cycle was hindered by ossification. Multicast routing, on the other hand, did not see wide-scale deployment because, in large part, it {\em anticipated} the Cycle by setting early goals that targeted future applications and future network scale.

\subsection{Unicast routing follows the Cycle}

Unicast routing protocols are used to build forwarding tables that guide network packets to one uniquely identified destination. Unicast routing is a fundamental function of networks. As such, it was the focus of research from the very early days of networking.

Table \ref{table:1} shows some (by no means all) important milestones in the evolution of deployed unicast routing protocols. 
The work by Prosser \cite{prosser1962routing} from 1962 represents work that laid the conceptual foundation for network routing.  The table also shows how, over the years, unicast routing has gone through many iterations of the Cycle. Each time a particular routing protocol is deployed experience is gained. This experience as well as the need to address increased network scale, cause the next iteration and a new version is deployed.  

Consider, for example, the deployment of the Exterior Gateway Protocol (EGP) which enabled unicast routing in a novel larger-scale Internet multi-domain structure. As the Cycle iterations continued, the Border Gateway Protocol (BGP) was developed a few years later and was 
``built on experience gained with EGP as defined in RFC 904 and EGP usage in the NSFNET Backbone as described in RFC 1092 and RFC 1093 "~\cite{lougheed1989rfc}; clear evidence that the Cycle was in operation.

The table also shows a vibrant landscape for unicast routing from the early 1960s. Things slowed down considerably by the early 2000s as ossification set in. BGP went from version 1 to version 4 in a span of 6 years, but nothing after that. Efforts to address security vulnerabilities in BGP have had mixed success because of the ``[lack of] a single 
centralized  authority that can mandate 
the  deployment"  \cite{goldberg2014taking}.

To summarize, the operation of the Cycle is evident in the history of innovation and deployment of unicast routing. The effect of ossification is also evident as further innovation in deployment stalls.

\begin{table*}[h]
\centering
\begin{tabular}{||l l l||} 
 \hline
 Year & Milestone & Comments \\ [0.5ex] 
 \hline\hline
 1962 & Prosser  \cite{prosser1962routing} & Conceptual Framework for Packet Routing  \\ 
 1969 & Distance Vector Routing \cite{heart1969} & First operational deployment of distributed routing \\
 1979 & Link State Routing \cite{mcquillan1979overview} & Better Scalability than Distance Vector Routing \\
 1982 & Exterior Gateway Protocol \cite{rosen1982rfc} & Enables simple multi-domain hierarchy\\
 1989-1995 & Border Gateway Protocol v1-v4 \cite{lougheed1989rfc,rekhter1995rfc} & Multi-domain policy Routing \\ 
 1999  & Multi-Protocol Label Switching  & Widely deployed in IP networks \\ 
 Early 2000s & Secure BGP \cite{kent2000secure,lynn2003secure} & Wide deployment lagging  \cite{goldberg2014taking}\\
 
 \hline
\end{tabular}
\caption{Unicast routing incrementally deployed before ossification}
\label{table:1}
\end{table*}

\subsection{Multicast routing anticipates the Cycle}

Now let's contrast the evolution of unicast routing with that of multicast routing. 
Multicasting is the act of sending data from a source to multiple destinations using a 
{\em single} transmit operation. Because of this, multicast can be highly efficient when applications call for it.  Multicasting can occur at several layers. Although sometimes it is not called out explicitly, it is used extensively at the application layer in group messaging (e.g., email, text) and live video streaming. For the most part, however, multicasting at the application layer is supported by multiple unicasts at the network layer, because multicasting at the network layer (i.e., multicast routing) is not widely deployed.

Multicast routing is typically deployed  today for niche applications that run on privately managed IP networks (see for example \cite{network2017national,IPTVArch}). There have been many discussions of why multicast is not deployed widely today in public networks (see for example \cite{diot2000deployment}). 

Milestones in the development of multicast routing are shown in Table \ref{table:2}. A good starting point is Dalal and Metcalfe's 1978 paper that, while aiming to propose the reverse path forwarding technique, summarizes most other options for deploying multicast routing. 

Probably the first proposal for deploying multicast in the Internet protocol stack as we know it today, was the 1984 paper by Aguilar in which he proposed using a form of multi-destination addressing, where the addresses of members of the group would be listed in each multicast packet. The technique had scalability limitations since one could only list up to 9 addresses in a single IP packet. Additionally, multicasting to this many destinations could easily be accomplished with separate unicast transmissions.
For these reasons Aguilar's proposal was not deployed.  

I argue that this was an {\em anticipation} of the Cycle.
This occurs when protocols are not deployed because they fail to satisfy some future requirement, even though they can be helpful in the present. In the case of Aguilar's multicast proposal, the requirement for highly scalable multicast routing
was not borne out of existing or even near term reality.
The Internet around that time had around 1000 hosts\footnote{https://www.zakon.org/robert/internet/timeline/}. Aguilar's mechanism would provide one order of magnitude scalability which would have been significant at the time. There was no urgent need for scalability beyond this potential order of magnitude savings. Furthermore, a multicast group exceeding 9 hosts could have easily been split into multiple smaller groups; something that was ultimately explored later \cite{kasera1997scalable,ammar1992improving}. A modestly scalable multicast routing deployment would have been extremely useful to get experience and provide a starting point for the Cycle's iterations.

Compare this with what happened in unicast routing, where the first distance vector protocol deployed had issues, including scalability limitations. Yet it served the purpose of getting things started. The main difference, of course, is the network had to have some form of unicast routing from the beginning, so we did not have the luxury of delaying deployment. 

As shown in Table \ref{table:2}, the late 1980s and early 1990s saw a progression of multicast routing protocols \footnote{The table only highlights routing developments and omits many important support mechanisms and protocols that were also needed to completely deploy multicast.}, each addressing the shortcomings of the one before or working to address a different set of constraints and goals.  This may look like the Service-Infrastructure Cycle in operation, except it is important to note that none of these protocols saw contemporaneous  {\em operational} deployment in support of applications or services beyond testbed environments such as the Mbone \cite{eriksson1994mbone}. 
Around 2005 there seemed to be a consensus around a protocol that enabled so-called {\em Single Source Multicast (SSM)} as one that met scalability and other requirements.  The problem is that by then the Internet had ossified and it was nearly impossible to deploy such a protocol, even though there were compelling arguments for its deployment.

\begin{table*}[h]
\centering
\begin{tabular}{||l l l||} 
 \hline
 Year & Milestone & Comments \\ [0.5ex] 
 \hline\hline
 1978 & Dalal and Metcalfe \cite{dalal1978reverse} & Conceptual Framework for multicast routing  \\ 
 1984 & Aguilar \cite{aguilar1984datagram} & First IP-based routing using multidestination addressing \\
 1988 & Distance Vector Multicast Routing (DVMRP) \cite{waitzman1988distance} & Deployed in the MBone experimental testbed \cite{Casner:1992:FII:142267.142338}\\
 1994 & Protocol Independent Multicast (PIM) \cite{deering1994architecture} & Support for sparse multicast groups \\
 2000 & Small-Group Multicast \cite{boivie2000small} & A modernized version of Aguilar's 1984 proposal \\
 2003 & Source Specific Multicast \cite{bhattacharyyaoverview} & Multicast support that finally addresses a real need \\ 
 
 \hline
\end{tabular}
\caption{Multicast routing does not get the benefit of deployment iterations before ossification}
\label{table:2}
\end{table*}

To summarize, multicast routing development never had a chance of ``riding" the Cycle. Early non-scalable ideas were rejected when scalability was not really needed. Experience with several generations of multicast routing protocols was gained only in testbed environments. When, finally, there appeared to be consensus on a workable multicast routing approach, it was no longer feasible to deploy it widely as ossification had already set in.

\subsection{Takeaways}

The takeaway message from this historical analysis is that before ossification, deployment success was achieved when functions had an opportunity to ride the Service-Infrastructure Cycle. Over multiple iterations, experience gained from deployment combined with motivation from new service or increased scale provided impetus for new and improved function deployment.

The two routing case studies illustrate how progress in support for novel services and in meeting demands imposed by increased scale requires the ability to ride the Service-Infrastructure Cycle. Delaying early deployment and waiting for mechanisms that satisfy long-term predicted goals had a detrimental impact on deployment success.

The case studies also show how ossification of infrastructure, when it occurs, can have a profound effect on the ability of the infrastructure to support novel services. In the unicast routing case, ossification made it difficult to widely deploy enhancements to BGP4. Multicast routing deployment was hindered because by the time there was agreement on deployable and practical techniques, ossification had set in. 

So what happens when the network infrastrcuture is unable to respond to new service or scale demands?  We discuss this question in the next section.

\section{From ManyNets to OneNet (and back again?)}

\label{manynets}

I now consider what happens when the ``unstoppable force" of continuously changing service demands and increasing scale meet the ``immovable object" that is the ossified Internet. I will first dwell on something that many take for granted, namely, the fact that today a common network infrastructure is being used to deliver communication services. I will then briefly discuss some of the new demands that the network infrastructure will need to address in the near future. Finally, I will argue that we are beginning to see evidence that the ``immovable object" is showing signs of yielding to the ``unstoppable force".

\subsection{The OneNet and its Consequences}

One of the most under-appreciated milestones in networking is the transition from a many networks (ManyNets) world to one where one common network (OneNet), the Internet, provides global connectivity for almost all services. 
The history of how  this transition occurred is well-documented \cite{leiner2009brief} and I will not repeat it here. Suffice it to say that economy of scale and ubiquitous connectivity, both of which were primary goals of {\em initial} networking efforts, are well-served with a common global network. 

The transition to OneNet was clearly in response to scale, global reach, and other demands on the network. As such, it fits within the general evolutionary paradigm of figure \ref{cycle}. The transition itself took time and many iterations of the Cycle. First all data communication, email, file download and some content services were consolidated. Networks and network technology that were so commonplace to be textbook material (see for example \cite{schwartz1987telecommunication}) like CCITT's X.25/X.75, IBM's SNA, Tymnet, and DECNET became of interest only as historical curiosities. Then, over time, voice and video services began to also be incorporated into the OneNet. 

The OneNet transition had important consequences.  The 
multiple-administrative-domains structure of the Internet today was conceived because it is not desirable to have a single entity manage a common network with global reach. Also a common network requires global agreement on network protocol standards which led to the OneNet agreement battles \cite{russell2006rough} ultimately leading to the adoption of the ARPANet architecture (a.k.a. the TCP/IP protocol suite) as the standard for global connectivity.

The OneNet world also has had strong implications for networking researchers.  For a number of years, there was little acceptance for work that could not be ultimately deployed in the Internet. 
The OneNet world also made it very difficult for researchers to validate their proposals at scale through experimentation. This led to the many efforts to build experimental networks: MBone \cite{eriksson1994mbone}, QBone \cite{teitelbaum1999internet2}, 6Bone \cite{fink20046bone}, PlanetLab \cite{chun2003planetlab}, and most recently GENI \cite{berman2014geni}.

For our purposes, however, the most important OneNet consequence was its inevitable ossification. This would not have happened had we stayed with ManyNets. {\em By its very nature, a ManyNets world cannot ossify,} since new demands can always be satisfied with new networks.

\subsection{New demands on network infrastructure}

Not surprisingly,
novel demands on the network infrastructure continue unabated. These demands need to be satisfied for the network infrastructure to maintain its usability. In the near term, the network infrastructure will need to handle:

\begin{itemize}
\item A significant increase in the number of connected devices with the advent of the Internet of Things (IoT). Estimates vary widely  with a rough consensus around a total of 14 billion additional connected devices on the Internet by 2022 \cite{spectrumconnecteddevices}.
\item The need for content providers to have increased control and accountability for the characteristics of network paths to their customers.
\item Unique service demands for emerging applications - low latency (gaming, real-time control, high-frequency trading), prioritization and preemption (critical communication), and high-bandwidth (high-quality video streaming).
\end{itemize}

\subsection{Fragmentation of network infrastructure}

So how is the networking infrastructure responding to these demands? Since the ossified OneNet cannot, it appears that we are slowly being ushered back into a fragmented ManyNets world. In this world, some (most?) services and content providers will use physically and/or virtually separate networks to reach their users. A shared infrastructure will continue to be available to reach services that don't find it feasible to provide a separated infrastructure or as backup. Individual users will possibly not be aware of this fragmentation. I discuss below some evidence of the emergence of this ManyNets world.

\paragraph{The Flattening of the Internet}

The earliest symptoms of Internet fragmentation were detected in the work by Gill et al \cite{gill2008flattening} in the form of a {\em flattening} of the Internet. The paper observes that ``large content providers are assembling their own wide-area networks." This was followed by a study by Labovitz et al \cite{labovitz2010internet}  that discovered that ``the majority of inter-domain traffic by volume now flows directly between large content providers, data center / CDNs and consumer networks." More recently Chiu et al \cite{chiu2015we} observe that ``Google connects directly to networks hosting more than 60\% of end-user prefixes, and that other large content providers have similar connectivity." 

\begin{figure*}[h]
    \centering
    \begin{subfigure}{0.42\textwidth}
        \centering
        \includegraphics[width=\linewidth]{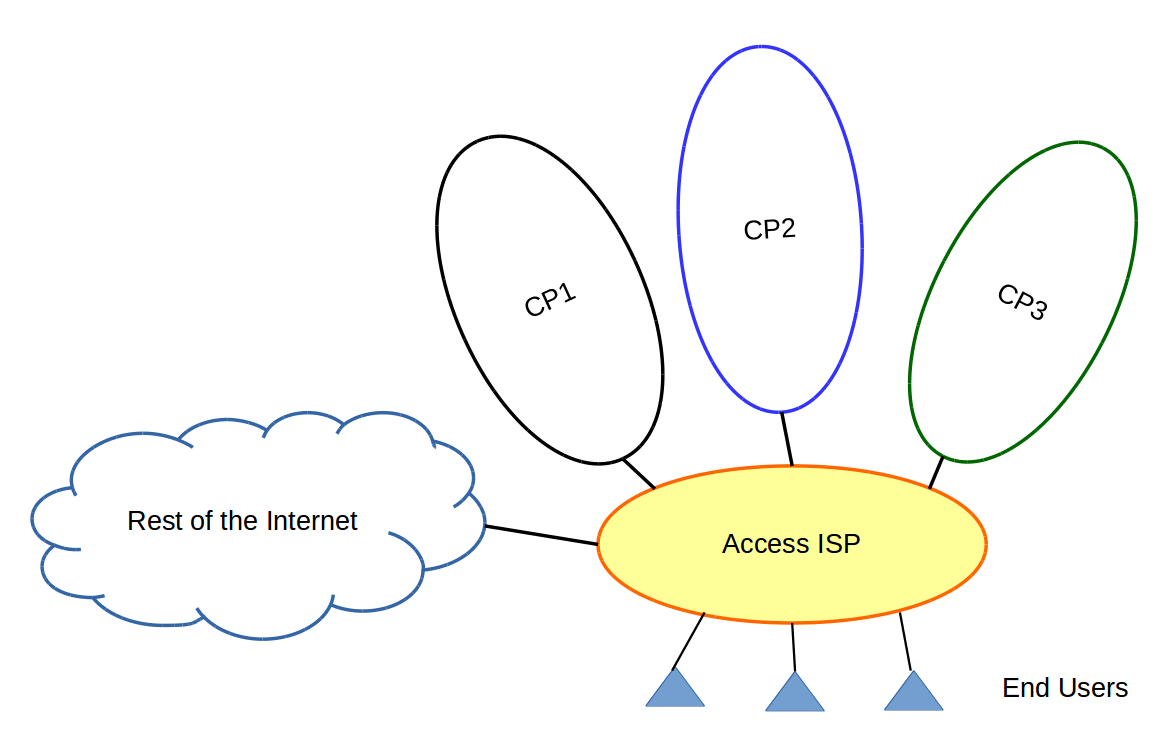}
  \caption{The flattening of the Internet}
  \label{FlatInternet}
    \end{subfigure}%
    ~
    \begin{subfigure}{0.42\textwidth}
        \centering
        \includegraphics[width=\linewidth]{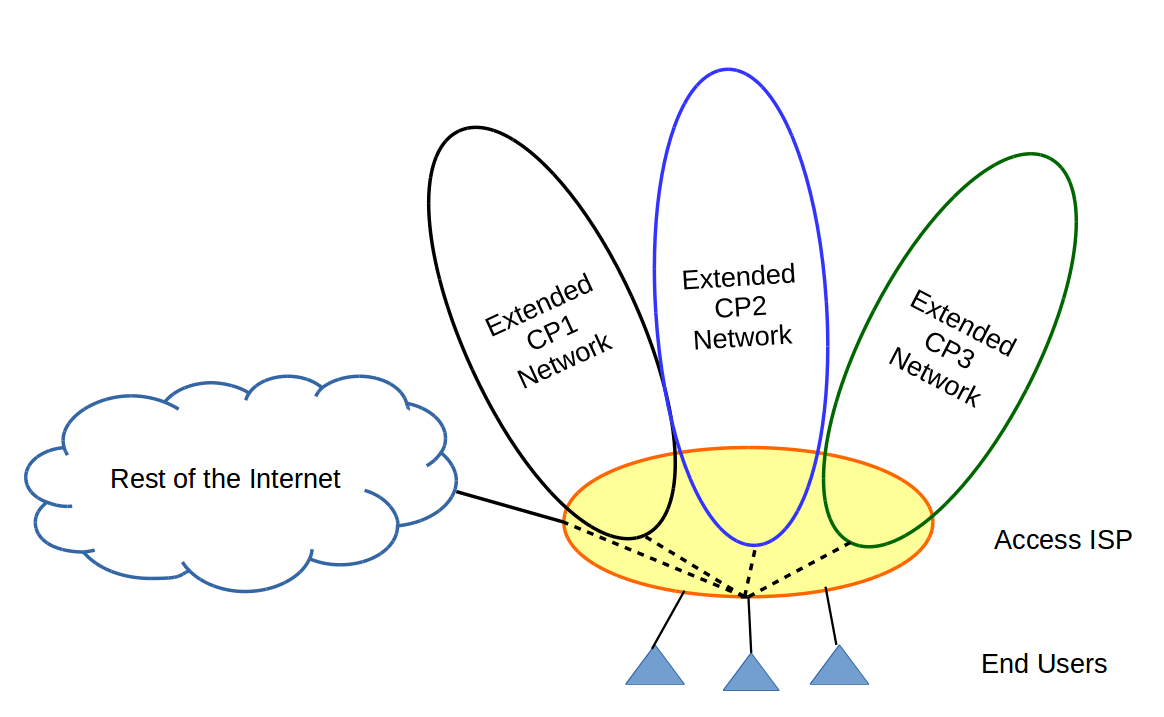}
  \caption{Zero-Hop Networking extends CP control into access ISPs}
 
  \label{zen}
        
    \end{subfigure}
    ~
    \caption{Flattening as a prelude to zero-hop fragmentation}
    \label{OneZero}
    \vskip -15pt
\end{figure*}

From a user's perspective, a flat Internet connectivity is as depicted in Figure \ref{FlatInternet}. Most of the Internet traffic travels over connections between a user's access network (access ISP) and a content provider (CP) network. 
The structure in Figure \ref{FlatInternet} represents, in principle, a common ubiquitous shared network providing global connectivity. In effect, however, for the majority of traffic this shared network exists to provide access to independently operated content provider network. 

The flattening of the  network infrastructure has been going on for some time. In itself, it does not necessarily represent the beginning of Internet fragmentation.  Recent trends, however, take this flattening a step further and are allowing content providers to {\em extend their control} into the shared access network itself. These include: 
\begin{itemize}
\item The placement of content provider servers (also known as appliances) within access networks \cite{netflixappliance,calder2013mapping}. These require access networks to provision specific bandwidth for access to these appliances, dedicating resources within access networks to specific content providers.

\item Specific frameworks and interfaces made available to content providers to enable collaboration and coordination between them and access networks (see for example \cite{ATTD2,ATTECOMP,frank2013pushing}). 

\end{itemize}

These recent trends represent, in effect, a fragmentation of the {\em access} network, the only part of the flat Internet shared among the CPs in Figure \ref{FlatInternet}. They are leading us into a  Zero-Hop Networking architecture \cite{ammar2017vision} as shown in Figure \ref{zen}. This further confirms that the initial flattening was, indeed, a first step to a more complete Internet fragmentation.

\paragraph{Emergence of Bypass Networks}

I define a {\em bypass network} as a network that operates and is administered separately from the OneNet global communication infrastructure. Such networks carry traffic with special requirements that cannot be accommodated by the global infrastructure for performance and/or capacity (or scale) reasons. By definition, such networks are a phenomenon of the OneNet world. Bypass networks can be {\em private,} built and operated for use by a specific enterprises or {\em public,} meaning anyone can subscribe, attach to and use the network.

Private bypass network examples include Google's own ``purpose-built network," \cite{jain2013b4,vahdat2015purpose} which was built because Google ``equipment vendors [were] slow in terms of delivering the capabilities requested by the company." This is another way of saying that the vendors (or is it the network standards the vendors were following?) were slow or incapable of following the Service-Infrastructure Cycle with Google's WAN demands being the driving force.

Another example of private bypass networks are the IPTV networks that cable television providers use to deliver digital content to users \cite{IPTVArch,network2017national}. These implement various technologies that are not deployed in the global infrastructure including IP Multicast to efficiently deliver digital content to cable modems. 

One can argue that private bypass networks do not necessarily represent a fragmentation of the global Internet infrastructure. Google's WAN and IPTV networks have limited accessibility and as such are not in competition with the OneNet.  

Motivated, however, by demands that the ossified OneNet cannot satisfy and, perhaps, encouraged by the feasibility of building separate networks demonstrated by private bypass networks, recent efforts have aimed to build global, publicly accessible, ``purpose-built" networks. These networks are true bypass networks because: 1) their traffic could have been carried on the global Internet, if only it were capable of delivering adequate service, and 2) are publicly available for subscription and use.

One of the earliest public bypass networks was the one built by Spread Networks \cite{spreadnetworks} in 2010 to support the low-latency requirements of high-frequency trading. The network boasts sub 13 ms round trip latency between New York and Chicago.

Another type of bypass network, that has seen recent significant development work and some global deployment, centers around the requirements of the Internet of Things (IoT). IoT has unique traffic requirements: traffic characterized often by the need for the sporadic exchange of short messages, the need to be power-efficient, and the expectation of very large scale deployments.  The belief that the current Internet infrastructure cannot in its current form handle these requirements and cannot be changed to meet them has led to the development of specialized network technologies and architectures, known collectively as Low-Power Wide-Area Networks (LPWANs)\cite{centenaro2016long}. LPWANs start with innovations in the access technologies, including the physical layer. Commercial deployments, however, such as SIGFOX \cite{zuniga2016sigfox} have included full networking protocol stacks and provide complete end-to-end service that does not include the use of Internet infrastructure.

Other examples of public bypass networks include {\em Haste} \cite{haste} and {\em FirstNet} \cite{firstnet}. Haste is a commercial bypass network that aims to provide very low latency for interactive gaming applications. It is a complete end-to-end solution that combines dedicated hardware as well as sufficient provisioning of resources to achieve its aims.  FirstNet \cite{firstnet} is a network built specifically with disaster and emergency first responder communication needs. Its functions include prioritization and preemption capabilities. 

Internet fragmentation is also evident in efforts to define 5G networks, a set of emerging conceptual architectures that aim to satisfy the demands of the future mobile and connected society \cite{shafi20175g}. Viewed within our Service-Infrastructure Cycle framework, 5G networks aim to provide the flexibility that will ensure the Cycle continues to  operate well into the future. A key enabler of this ``lubrication" of the Cycle is the idea of 5G slicing. In the 5G slicing proposal \cite{5GHuwawei} the end-to-end network is sliced into networks with three different sets of capabilities: 1) A high throughput slice, 2) an ultra low latency slice, and 3) a large scale connectivity slice (e.g., for IoT devices). While all of this is still on the drawing board for now, it clearly points to a mindset where the future networking needs are no longer satisfied by a common global infrastructure.

\section{Concluding Remarks}

In the 1970s, a group of networking researchers and engineers set out to deploy a communication infrastructure that would act as 
``utilities [that] will offer a variety of information services and transactions, such as retrieval from multiple independent databases, messaging, electronic mail, conferencing, banking, tele-shopping and interest matching."
Meanwhile around the same time another group of communication engineers were beginning to deploy another communication network. That network, while motivated by a grand vision from the decade before of providing ``Man-Machine Symbiosis" and of ``The Computer as a Communication Device", started with the simple goal of providing computer-to-computer connectivity and resource sharing.

If you guessed that the first communication infrastructure was the precursor of today's Internet, you are mistaken. It was the vision of what were called then Videotex networks \cite{ball1980videotex}. The second network -- now you know -- was indeed the Internet's direct ancestor \cite{leiner2009brief}, delivering the beginnings of a vision laid out by J. C. R. Licklider in the 1960s\cite{JCR}.

Today the first network, Videotex, does not exist because it attempted to meet all its goals in one shot \cite{MICHAELNOLL198599}.  The second became the Internet and provides all the services envisioned by the first network and then some.  I have argued here that the success of the Internet today in delivering services should be attributed to a modest start that was followed by iterations of the Service-Infrastructure Cycle. Similarly the failure of the Videotex network was not because of its vision, but rather because it anticipated the Cycle and tried to accomplish too much, too fast.

I have also argued and provided some evidence that the ossification of the Internet infrastructure along with continued new demands are requiring the fragmentation of the infrastructure so that the Service-Infrastructure Cycle can continue to operate -- {\em from one, many}.

We are left with the question: {\em So what if the communication infrastructure is fragmenting?}
Overall, it is not a bad thing and we may not have any choice in the matter.  I do, however, have a couple of concerns.

My first concern is not about the fragmentation itself but that, despite the re-emergence of the ManyNets world, the networking research community might continue to expend its energy to further the illusion of a single global networking infrastructure. For a while after infrastructure ossification was observed, there was a search for either a fix that would de-ossify the Internet  (through virtualization, for example \cite{turner2005diversifying}), or a single replacement for the Internet that would be immune from ossification by design \cite{pan2011survey}. While these efforts have had significant intellectual impact, they have yet to have a perceptible impact on the deployed infrastructure. The idea of a single unified networking infrastructure was perhaps worth fighting for and saving. However, this battle is lost. We should, therefore, embrace the ManyNets world and the research agenda that emerges from it. 

In fact, the emergence of the ManyNets world will be a boon to networking research. It is a fertile ground for innovation as increased scale, with no end in sight, and new services continue to drive network evolution.

The second concern I have is that, while it lasted, the OneNet has served well in providing cost-efficient networked services. Perhaps, more importantly, it has enabled significant innovation over the years by providing a low cost of entry to new services. The challenge for the networking research community is how to continue these benefits in the emerging ManyNets world.

\section{Acknowledgments}
Thanks to Jorg Liebeherr for giving me the time and space at the University of Toronto to work on this article during summer of 2017. The many conversations we had on this topic and his feedback on an early draft helped shape some of the arguments. Marwan Fayed and Ellen Zegura also provided valuable feedback on early drafts. Ken Calvert, Raouf Boutaba and Yashar Ganjali provided valuable feedback through their contributions to the panel at the 2017 IEEE ICNP conference that debated the arguments in this article and considered its implications on the networking research community.
%
\bibliographystyle{abbrv}
\bibliography{HistPersp}  

\end{document}